\documentclass[%
 reprint,
 amsmath,amssymb,
 aps,
]{revtex4-2}

\usepackage{graphicx}
\usepackage{dcolumn}
\usepackage{bm}
\usepackage{braket}

\begin{document}

\preprint{APS/123-QED}

\title{Correlated oscillations in Kerr parametric oscillators with tunable effective coupling}

\author{T.~Yamaji$^{1, 2}$}
\thanks{Corresponding author: tyamaji@nec.com}
\author{S.~Masuda$^{2, 3}$}
\author{A.~Yamaguchi$^{1, 2}$}
\author{T.~Satoh$^{1, 2}$}
\author{A.~Morioka$^{1, 2}$}
\author{Y.~Igarashi$^{1, 2}$}
\author{M.~Shirane$^{1, 2}$}
\author{T.~Yamamoto$^{1, 2}$}

\affiliation{$^1$Secure System Platform Research Laboratories, NEC Corporation, Kawasaki, Kanagawa 211-0011, Japan}
\affiliation{$^2$NEC-AIST Quantum Technology Cooperative Research Laboratory, National Institute of Advanced Industrial Science and Technology (AIST), Tsukuba, Ibaraki 305-8568, Japan}
\affiliation{$^3$Research Center for Emerging Computing Technologies (RCECT), National Institute of Advanced Industrial Science and Technology (AIST), Tsukuba, Ibaraki 305-8568, Japan}
\date{\today}

\begin{abstract}
We study simultaneous parametric oscillations in a system composed of two distributed-element-circuit Josephson parametric oscillators in the single-photon Kerr regime coupled via a static capacitance.
The energy of the system is described by a two-bit Ising Hamiltonian with an effective coupling whose amplitude and sign depend on the relative phase between parametric pumps.
We demonstrate that the binary phases of the parametric oscillations are correlated with each other, and that the parity and strength of the correlation can be controlled by adjusting the pump phase. 
The observed correlation is reproduced in our simulation taking pure dephasing into account.
The present result demonstrates the tunability of the Hamiltonian parameters by the phase of external microwave, which can be used in the Ising machine hardware composed of the KPO network.

\end{abstract}

\maketitle

\section{Introduction}
A quantum annealer is a system consisting of a network of qubits, designed to perform quantum annealing, which is a method searching for global minima of an Ising Hamiltonian encoded on the network \cite{Kadowaki1998}. 
A wide range of optimization problems can be formulated as combinatorial optimization problems whose cost functions are expressed as an Ising Hamiltonian \cite{Lucas2014}. 
In the hope of solving large-scale industrial and social optimization problems in a reasonable time, quantum and classical annealers have been developed using a variety of architectures. 
D-Wave Systems has developed commercial quantum annealers composed of superconducting flux qubits \cite{Johnson2011, King2021, Raymond2022, King2022, King2022a}. 
A novel annealer called a coherent Ising machine has been developed using optical systems \cite{Wang2013, Marandi2014, McMahon2016, Inagaki2016}.
Classical annealers have also been developed using conventional devices such as FPGAs and GPUs \cite{Yamaoka2016, Matsubara2018, Goto2019b}. 

A Kerr Parametric Oscillator (KPO) has been recently proposed as a new candidate for a building block of a quantum annealer \cite{Goto2016, Nigg2017, Puri2017a, Zhao2018, Goto2019a, Onodera2020, Goto2020, Kewming2020, Kanao2021}. 
A parametric oscillator is a nonlinear resonator whose parameters can be modulated by an external force called a parametric pump, and exhibits binary self-oscillating states with a phase of either $0$ or $\pi$ \cite{Nayfeh1995, Dykman1998, Strogatz2000}.
A parametric oscillator in the single-photon Kerr regime is called KPO, where the nonlinearity such as the Kerr effect is stronger than dissipation \cite{Kirchmair2013}.
A KPO also has various applications in the field of quantum information such as deterministic generation of Schr\"{o}dinger cat state \cite{Goto2016, Puri2017, Goto2019},
a qubit for quantum logic gate \cite{Goto2016a, Puri2017, Grimm2020, Frattini2022}, study on the quantum chaos \cite{Goto2021}.

The KPOs have been experimentally realized by using trapped ions \cite{Ding2017} and Josephson parametric oscillators (JPOs) \cite{Wang2019, Grimm2020, Yamaji2022, Frattini2022}.
However, realization of a KPO network remains elusive. 
The KPO-network implementation requires tunable bit-to-bit coupling \cite{Masuda2022} as used in the flux-qubit-based architecture, where the tunable coupling is realized by tunable couplers made of flux qubits \cite{Weber2017}, and it is desirable that the sign and amplitude of the coupling can be adjusted independently of variations in device fabrication.  

Here, we design and fabricate a device composed of two JPOs in the single-photon Kerr regime coupled via a static capacitance, which correlates parametric oscillations of the JPOs.
We measured simultaneous parametric oscillations in a steady state, and observed that the correlation between simultaneous parametric oscillations can be controlled {\it in situ} by varying the relative phase between parametric pumps, as proposed in Ref.~\citenum{Masuda2022}, despite the use of a static capacitance. 
The experimental result is well consistent with numerical simulations, which solve the master equation with pure dephasing taken into account.
This is a demonstration of control of the effective coupling between KPOs, which corresponds to the coupling between spins in the encoded Ising Hamiltonian.
The tunability of the coupling is indispensable for quantum annealing since optimization problems are generally mapped to Ising Hamiltonians consisting of spins coupled with various coupling strengths in both polarities (ferromagnetic and antiferromagnetic).
This work paves the way for the implementation of a KPO-based quantum annealer.

\section{Device}

\begin{figure}
	\includegraphics[width=\linewidth]{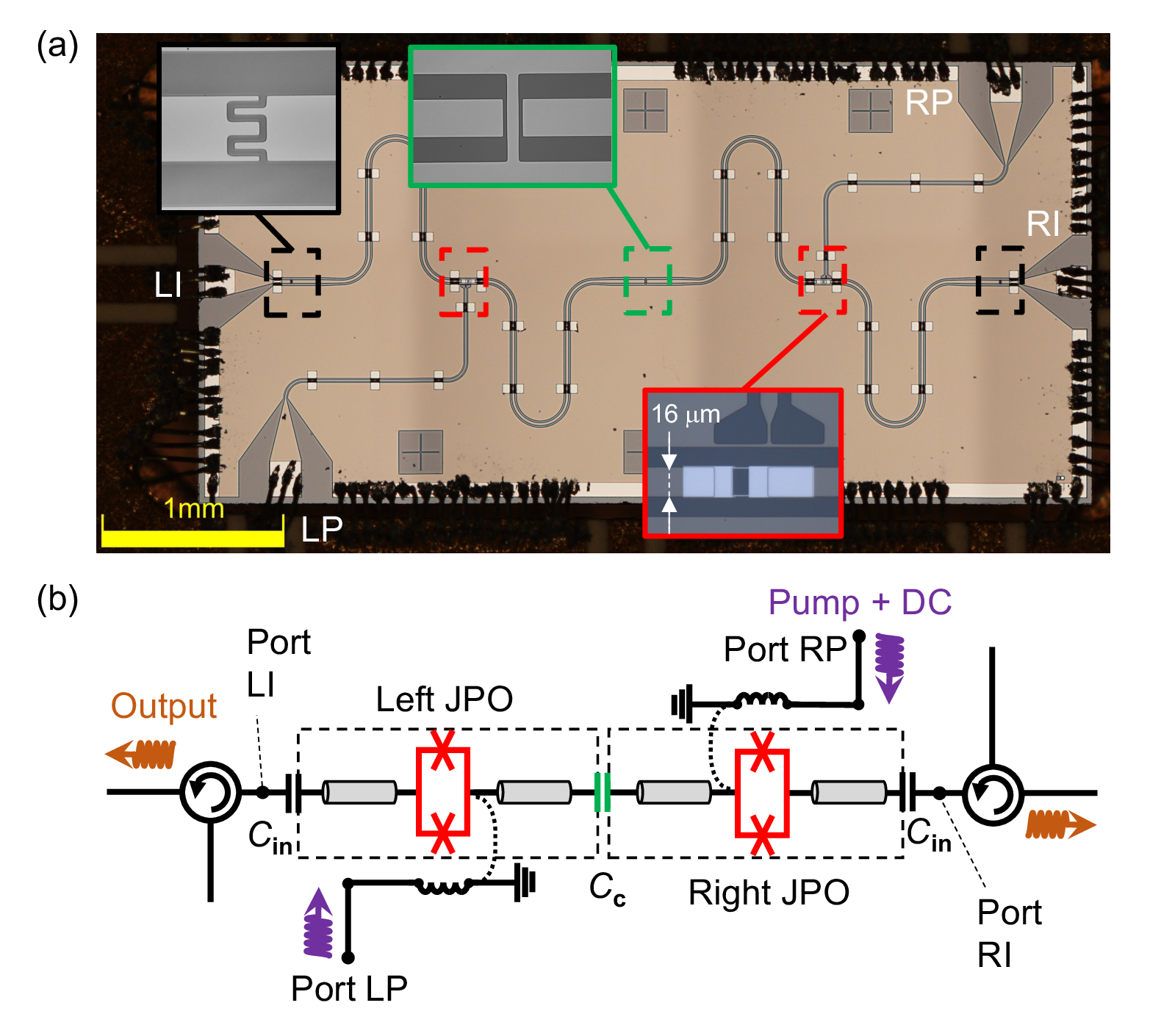}
	\caption{
	Device chip studied in this paper. 
	(a) Optical image. 
	The red, black, and green insets show the magnified views around the SQUID, the I/O capacitor $C_{\rm in}$, and the coupling capacitor $C_{\rm c}$, respectively. 
	(b) Schematic circuit diagram. 
	The grey tubes surrounding the SQUIDs represent coplanar waveguides (CPWs).
	Circulators are used to separate input and output microwaves for each JPOs. 
	The four ports of the device chip are labeled as LI, LP, RI, and RP.
}
	\label{fig:sample}
\end{figure}
 
Figure~\ref{fig:sample} shows the device chip studied in this paper.
The device has two JPOs on the left and right sides, which are labeled L and R, respectively.
Each JPO consists of a half-wavelength, 4.614-mm long, CPW resonator, whose characteristic impedance and phase velocity are designed to be 49.8~$\Omega$ and $0.398c$, respectively, where $c$ is the speed of light. 
The inner ends of the resonators, located at the center of the device, form a coupling capacitor $C_{\rm c}$ of $0.6$~fF, which is designed using electromagnetic field simulation with a structure larger than the fabrication uncertainty. The outer end of each resonator is connected to a feed line via an input and output (I/O) capacitor $C_{\rm in}$ of $3.4$~fF [LI and RI ports shown in Fig.~\ref{fig:sample}(b)].
Each resonator is interrupted by a symmetric DC-SQUID at the center of the resonator.
The critical current $I_{\rm c}$ of each Josephson junction in the SQUID is estimated to be $1.17~{\rm \mu A }$ from the maximum resonance frequency of the resonator, 11.95~GHz. 
The estimation is consistent with the room-temperature resistance of a test structure on the same chip.
Each SQUID is inductively coupled to a pump line, on which a DC current to control the resonance frequency and a pump microwave to induce parametric oscillations are applied  (LP and RP ports). 
The resonators, the feed lines, and the pump lines are equipped with lithographically patterned airbridges to suppress parasitic slotline modes of the CPWs and to reduce AC/DC crosstalk between the JPOs (See Appendix A for details). 
The device chip is stored in a magnetic shield and cooled below 10 mK in a dilution refrigerator. 
Hereinafter, all the input/output powers of the JPOs are specified at the relevant ports on the device chip. 

\section{Hamiltonian and parameters}
In the present paper, we consider the case where both the JPOs are modulated at the same pump frequency~$\omega_{\rm p}$. 
Under the rotating wave approximation (RWA), the Hamiltonian of the coupled-JPO system in a frame rotating at half the pump frequency is given by \cite{Goto2018}
\begin{eqnarray}
\mathcal{H}/\hbar&=&\sum_{i={\rm L,R}}\left[\frac{K_i}{2}a_i^{\dag 2} a_i^2 + \Delta_i a_i^\dag a_i + \frac{p_i}{2}\left(a_i^{\dag 2} + a_i^2\right) \right] \nonumber \\
&&+g \left(e^{-{\rm i}\theta_{\rm p}/2}a_{\scalebox{0.5}{\rm L}}^\dag a_{\scalebox{0.5}{\rm R}} + e^{{\rm i}\theta_{\rm p}/2}a_{\scalebox{0.5}{\rm L}} a_{\scalebox{0.5}{\rm R}}^\dag \right), \label{eq:H}
\end{eqnarray}
where $K_i\ (<0)$ is the Kerr nonlinearity; $\Delta_i \equiv  \omega_{{\rm r}i} -\omega_{\rm p}/2$ is the detuning of the oscillation frequency $\omega_{\rm p} / 2$ from the resonance frequency $\omega_{{\rm r}i}$; $p_i\ (>0)$ is the pump amplitude; 
$\theta_{\rm p}$ is the phase difference of the pumps applied to the JPOs (the R pump phase relative to the L pump phase); $a_i$ is the annihilation operator for the JPOs, and $i=$L/R represents the index of the JPOs hereafter. 
It is important to note that the phase factor in the coupling term originates from unitary transformation  $a_{\rm \scalebox{0.5}{R}} \to e^{-{\rm i}\theta_{\rm p}/2} a_{\rm \scalebox{0.5}{R}}$, which absorbs the phase difference between the parametric pumps and makes both the pump amplitudes $p_i$ real.
The ground state without the pump and the coupling is the vacuum state $\ket{0}_{\rm \scalebox{0.5}{L}}\ket{0}_{\rm \scalebox{0.5}{R}}$, which corresponds to the maximum energy state in a rotating frame due to the negative Kerr nonlinearity.

While the oscillation states of each JPO can be approximated as coherent states $\ket{\pm \alpha_i}_i$ with an amplitude of $\alpha_i \simeq\sqrt{(p_i+\Delta_i)/\left|K_i\right|}$ by considering the coupling as perturbation, those of the entire system can be expressed as a tensor product of the coherent states,  $\ket{s_{\scalebox{0.5}{L}}\alpha_{\rm \scalebox{0.5}{L}}}_{\rm \scalebox{0.5}{\rm L}}\ket{s_{\scalebox{0.5}{R}}\alpha_{\rm \scalebox{0.5}{R}}}_{\rm \scalebox{0.5}{\rm R}}$, where $s_i=+1\ (-1)$ represents the Ising spin corresponding to the oscillation state with a phase of 0~($\pi$) \cite{Goto2018}.
These oscillation states can be differentiated by simultaneously measuring the relative phase of parametric oscillations.
The eigenenergies of the oscillation states are obtained by replacing the annihilation operator $a_i$ in Eq.~(\ref{eq:H}) with $\alpha_i$,
\begin{equation}
E/\hbar=\sum_{i={\rm L,R}}\left[\frac{K_i}{2}\alpha_i^4 + \Delta_i \alpha_i^2 + p_i \alpha_i^2 \right] - \left[-Js_{\scalebox{0.5}{\rm L}}s_{\scalebox{0.5}{\rm R}}\right], \label{eq:Ising}
\end{equation}
where $J=2{\rm cos}(\theta_{\rm p}/2)g\alpha_{\scalebox{0.5}{\rm L}}\alpha_{\scalebox{0.5}{\rm R}}$ is the effective coupling.
The eigenenergy contains the Ising energy 
$E_{\rm Ising}=-J s_{\scalebox{0.5}{\rm L}}s_{\scalebox{0.5}{\rm R}}$ 
with a coupling constant controllable by the pump phase. 
Since the negative Kerr nonlinearity favors the state with the highest eigenenergy in the rotating frame, the initial state evolves to the oscillation state minimizing the Ising energy as the pump is gradually applied, yielding a solution to the Ising Hamiltonian encoded on the device.

\begin{figure}
	\includegraphics[width=1.0\linewidth]{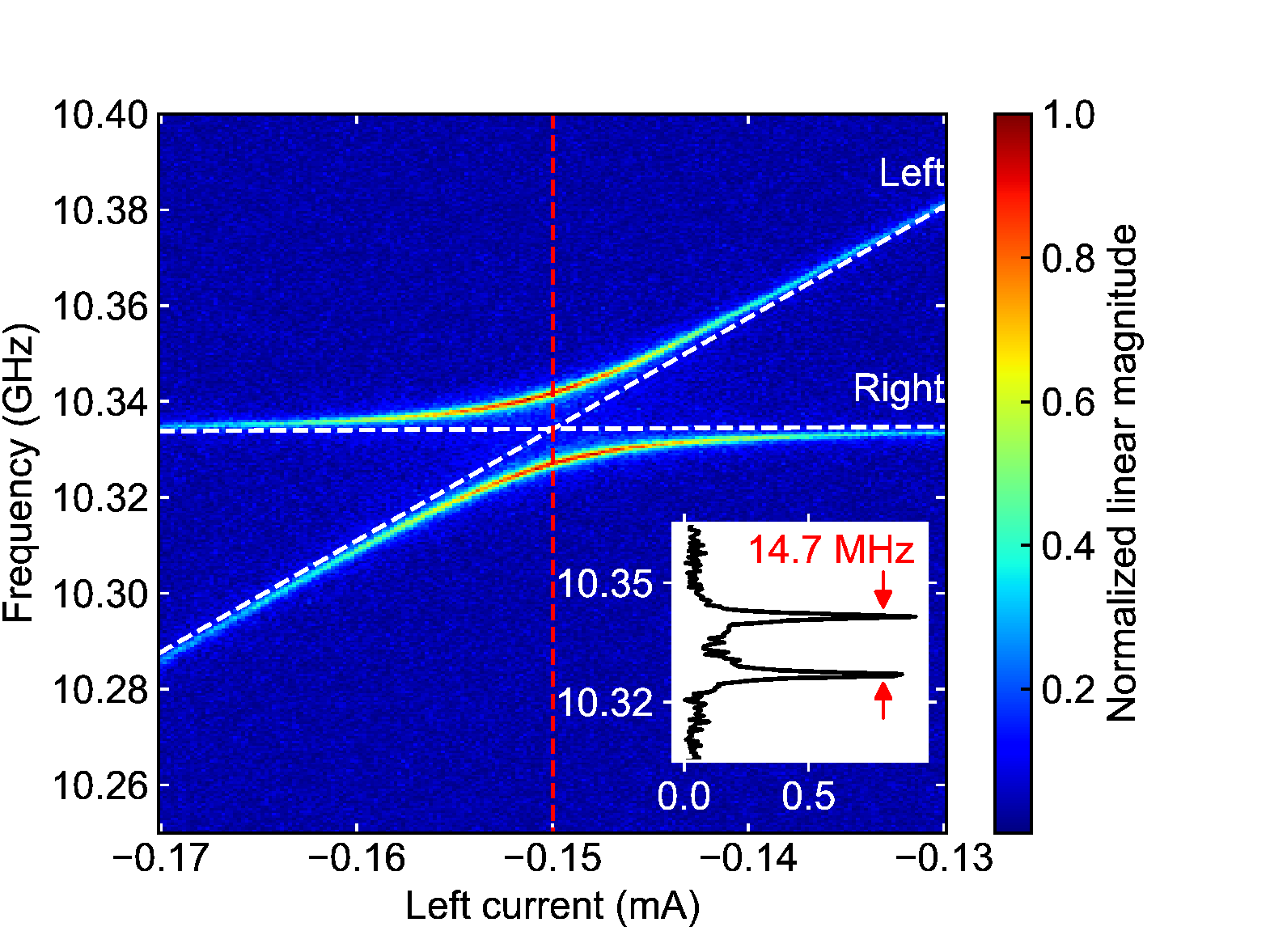}
	\caption{
	DC-flux-bias dependence of the transmission coefficient from the LI to RI ports of the device. 
	The coefficient is measured in the low probe-power limit ($-150$~dBm on the LI port). 
	The horizontal axis shows the DC current applied to the LP port. 
	The vertical axis shows the frequency of the probe microwave. 
	The color scale shows the normalized magnitude of the transmission coefficient.
	The white dashed lines represent the uncoupled resonance frequencies of the JPOs. 
	The inset shows the cross section along the red dashed line, where the horizontal and vertical axes of the inset show the magnitude of transmission coefficient and the probe frequency, respectively.}
	\label{fig:anti-crossing}
\end{figure}

In order to fix an operating point for the measurement of the simultaneous parametric oscillations, we first measured the dc-flux-bias dependence of the resonance frequencies of the two JPOs.
Figure~\ref{fig:anti-crossing} shows the microwave transmission from the LI to RI ports as a function of the DC current applied to the LP port.
The two peaks correspond to the resonance frequencies of the two JPOs.
The transmission coefficient becomes larger when the uncoupled resonance frequencies of the L(R) JPOs $\omega_{\rm r\scalebox{0.5}{L(R)}}$ are close to each other. 
They show the clear avoided level crossing, whose minimum frequency splitting of 14.7~MHz is twice the coupling constant $g$ between the JPOs induced by the coupling capacitance.
The uncoupled resonance frequency is $\omega_{\rm r\scalebox{0.5}{L(R)}}/2\pi=\omega_{\rm r}/2\pi=10.3342$~GHz.
We note that $\omega_{\rm r\scalebox{0.5}{R}}$ weakly depends on the DC current applied to the LP port due to 0.8\% crosstalk between the JPOs. 

The external and internal photon loss rates of the JPOs at $\omega_{\rm r}$ are estimated to be $\kappa_{\rm e\scalebox{0.5}{L(R)}}/2\pi=0.82(0.63)$~MHz and $\kappa_{\rm i\scalebox{0.5}{L(R)}}/2\pi=0.18(0.23)$~MHz, respectively, from reflection coefficient measurement when the other JPO is detuned. 
We note that the measured internal loss rates may include contribution from pure dephasing: the actual internal loss rate $\kappa_{\rm i\scalebox{0.5}{L(R)}}^\ast$ is related to $\kappa_{\rm i\scalebox{0.5}{L(R)}}$ as $\kappa_{\rm i\scalebox{0.5}{L(R)}}^\ast = \kappa_{\rm i\scalebox{0.5}{L(R)}} - 2 \gamma_{\rm \scalebox{0.5}{L(R)}}$, where $\gamma_{\rm \scalebox{0.5}{L(R)}}$ is the pure dephasing rate of the L(R) JPO. 
The Kerr nonlinearities are also evaluated to be $K_{\rm L(R)}/2\pi=-10.4(-10.3)$~MHz from the resonance transition at $\omega_{\rm r} +K$ in the two-tone spectroscopy \cite{Yamaji2022}.
The JPOs are well in the single-photon Kerr regime, $\left|K_i\right|\sim 10\kappa_{{\rm a}i}$, where $\kappa_{{\rm a}i}\equiv\kappa_{{\rm e}i}+\kappa_{{\rm i}i}$.
In the following experiments we fix the DC flux bias shown by the vertical dashed red line in Fig.~\ref{fig:anti-crossing}, where $\omega_{\rm r\scalebox{0.5}{L(R)}} = \omega_{\rm r}$. 

\begin{figure}
	\includegraphics[width=1.0 \linewidth]{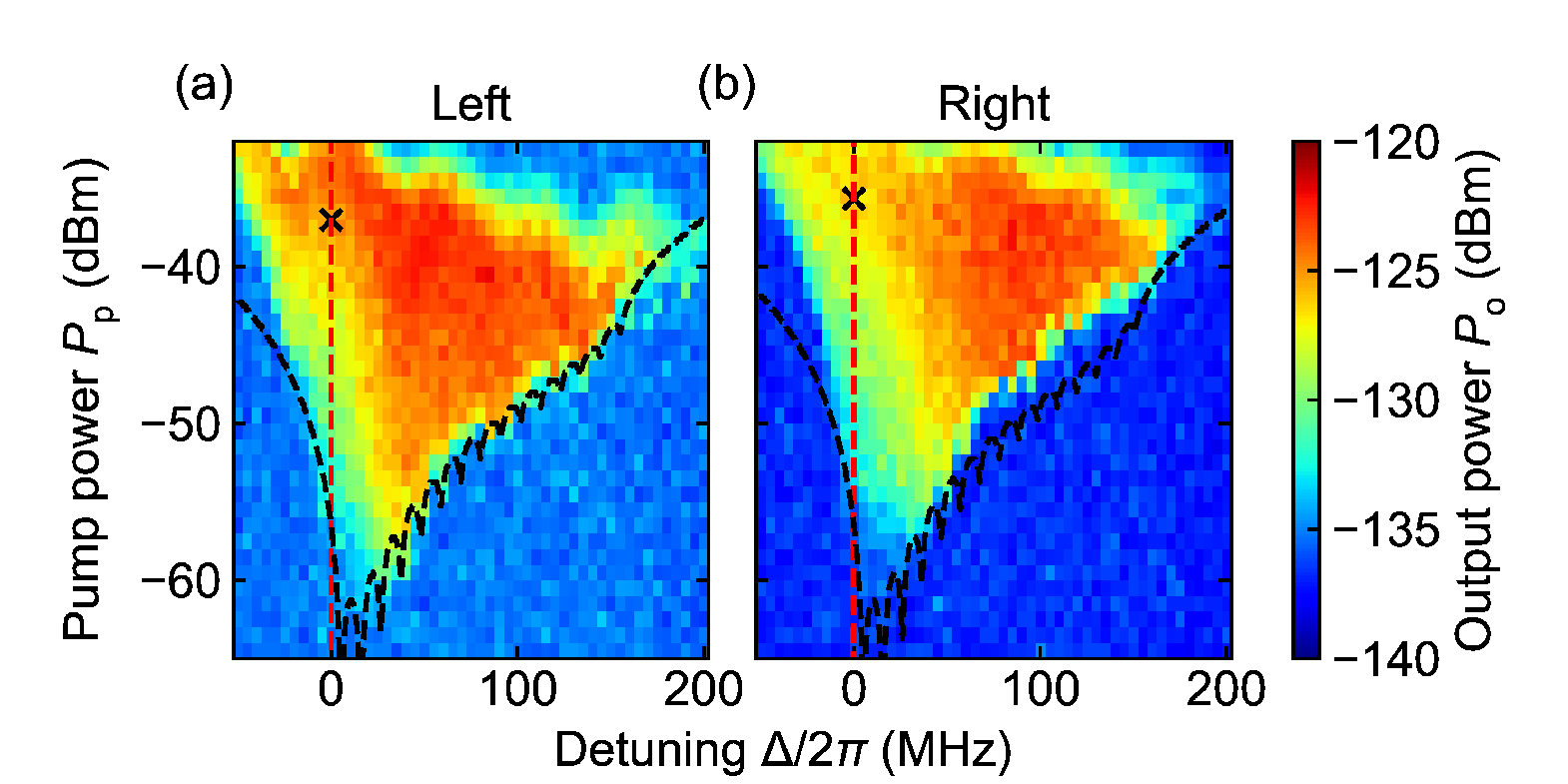}
	\caption{
	Continuous-wave (CW) parametric oscillations of the L JPO (a) and the R JPO (b). 
	The horizontal and vertical axes show the detuning $\Delta=\omega_{\rm r}-\omega_{\rm p}/2$ and the pump power $P_{\rm p}$, respectively. 
	The color scale shows the output power of the parametric oscillation $P_{\rm o}$. 
	The static resonance frequency is shown by the red dashed line. 
	Black crosses show the operating point in the time-domain measurements. 
	The black dashed lines show the calculated pump power where the mean photon number in the JPOs is unity [$P_{\rm o\scalebox{0.5}{L(R)}}=-135(-136)\ {\rm dBm}$].
	}
	\label{fig:para_ex}
\end{figure}

Figure~\ref{fig:para_ex} shows continuous-wave~(CW) parametric oscillations of each JPOs.
The parametric oscillations are induced by individually applying CW pump microwaves at $\omega_{\rm p}$ to each JPO, while the pump for the other JPO is turned off.
The output power $P_{\rm o}$ at $\omega_{\rm p}/2$ are measured by a spectrum analyzer as a function of $\omega_{\rm p}$ and the power of the pump $P_{\rm p}$.
For both the JPOs, as we increase $P_{\rm p}$ above a certain power which depends on $\omega_{\rm p}$, we observe the output power indicating parametric oscillations. 
To understand this behavior, we calculated the mean photon number $n$ in the JPO based on the analytical formula for the steady-state \cite{Bartolo2016}, and plotted $P_{\rm p}$ corresponding to $n = 1$ by the black curves in Fig.~\ref{fig:para_ex}. 
All the parameters in the calculation are determined from the independent measurements. 
The calculation well reproduces the overall trend in the experiment. 
We note that the periodic structure in the calculation on the positive detuning side is not clearly observed in the present experiment, which is probably due to the insufficient resolution of the frequency step. 
The structure has an interval of $|K|$, and is also pointed out in Ref.~\cite{Wang2019}. 

In the following time-domain measurements, we set the detuning $\Delta \equiv \omega_{\rm r} - \omega_{\rm p}/2$ to be zero, namely $\omega_{\rm p}/4\pi =\omega_{\rm r}/2\pi= 10.3342$~GHz, for simplicity.
Note that under this condition, the vacuum state is not the highest energy state in the rotating frame because of the coupling term in the Hamiltonian. 
We set pump power $P_{\rm p \scalebox{0.5}{L(R)}}=-37\ (-36)$~dBm (the black crosses in Fig.~\ref{fig:para_ex}), 
because the output power is stably high and similar for the two JPOs around the operating point. 
The output power at the point is $P_{\rm o \scalebox{0.5}{L(R)}}=-126\ (-128)$~dBm, which corresponds to the amplitude of the coherent state of $\alpha_{\rm \scalebox{0.5}{L(R)}}=\sqrt{P_{\rm o \scalebox{0.5}{L(R)}}/(\hbar\omega_{\rm r}\kappa_{\rm e\scalebox{0.5}{L(R)}})}=2.8\ (2.5)$. 

\begin{figure*}
	\includegraphics[width=1.0 \linewidth]{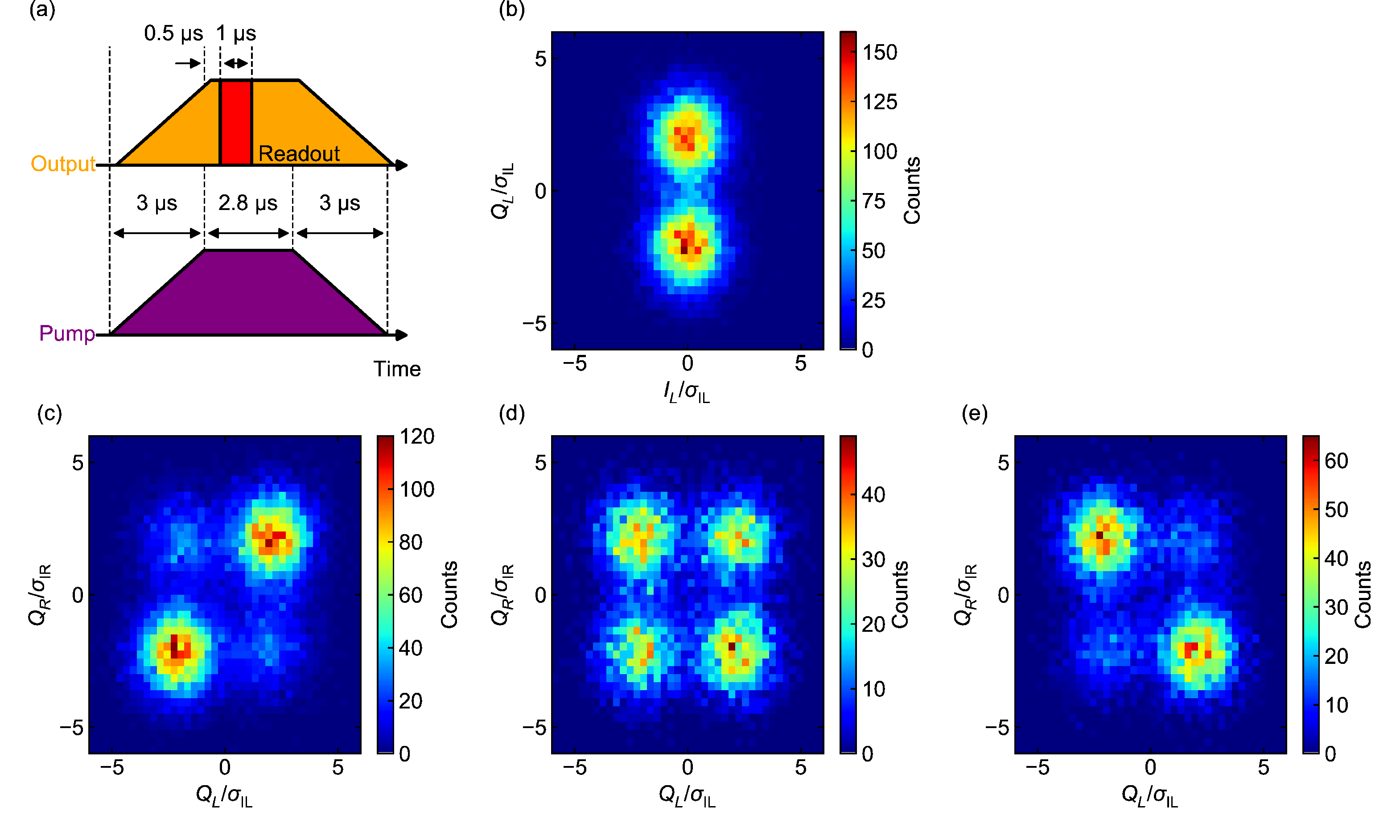}
	\caption{
	Time-domain measurement. 
	(a) Pulse sequence of the pump and the induced parametric oscillation output. 
	The pump power and the signal frequency are fixed to $P_{\rm pL(R)}=-37(-36)$~dBm and $\omega_{\rm p}/4\pi=\omega_{\rm r}/2\pi=10.3342$~GHz, respectively. 
	(b) The $IQ$-plane histogram of the output from the L JPO. 
	The translational and angular offsets of the output is calibrated by tuning the center-of-mass of the data to the origin and aligning the two peaks parallel to the $Q$ axis. 
	(c) The histogram of the $Q$ values of the outputs from the JPOs. 
	 The histograms shown in (b-c) is obtained by integrating $2\times 10^4$ shots at $\theta_{\rm p}=0$.
	 (d-e)  The histograms similar to (c), which integrates $1\times 10^4$ shots at $\theta_{\rm p}=\pi$ and $2\pi$, respectively.   
	 }
	\label{fig:correlation}
\end{figure*}

\section{Correlation in parametric oscillations}
In order to investigate the effect of the coupling on the oscillation phases of the JPOs, we performed the time-domain measurement by simultaneously applying pulsed pumps to the JPOs at the operating point shown in Fig.~\ref{fig:para_ex} with the pulse sequence shown in Fig.~\ref{fig:correlation}(a).
We used the pulsed pump to reset the oscillation state and repeat the measurement to take statistical average.
The pulsed pumps are trapezoidal with a slope of $3\ {\rm \mu s}$ and a plateau of $2.8\ {\rm \mu s}$. 
The minimum energy gap during the pulse sequence is estimated to be $\mathcal{O}\left(|K|\right)$ \cite{Puri2017a}, and the slope is sufficiently longer than $|K|^{-1}=15$~ns in order to suppress unwanted nonadiabatic transitions.
In addition, the pump slope is much longer than $1/\kappa_{{\rm a}i}=0.2~{\rm \mu s}$ to evaluate coupling tunability in a steady state insensitive to the pump slope (See also Appendix~\ref{sup:adiabatic}).
The readout is delayed by $0.5\ {\rm \mu s}$ from the start of the pump plateau to wait for the saturation of the JPOs.
The output signals are recorded in the heterodyne measurement with an integration time of $1~{\rm \mu s}$ (See Appendix~\ref{sup:meas-setup} for details), where the integration time is determined by the signal-to-noise ratio such that the peaks of the $Q$ amplitudes of the oscillation states are more than $2\sigma$ away from the origin.

Figure~\ref{fig:correlation}(b) shows the histogram of the output signal from the L JPO plotted in the in-phase and quadrature $(IQ)$ plane. The distribution is obtained by integrating $2 \times 10^4$ shots in the condition that $\theta_{\rm p}=0$.
The histogram has two equally distributed peaks with an equal amplitude and well-defined phases shifted by $\pi$, which correspond to the coherent states, $\ket{\pm \alpha_{\rm \scalebox{0.5}{L}}}_{\rm \scalebox{0.5}{L}}$.
Because possible leakage from the other JPO can break each peak in two, its absence shows that the contribution of the leakage is small and negligible in the measurement.
The occurrence probabilities of the coherent states, $\ket{\pm \alpha_{\rm \scalebox{0.5}{L}}}_{\rm \scalebox{0.5}{L}}$, are the same because they are degenerate.

Figure~\ref{fig:correlation}(c) shows the histogram of the $Q$ amplitudes of the two JPOs at $\theta_{\rm p}=0$.
Although the state of each JPO is randomly determined, the probability of the same-phase configuration $\ket{\pm\alpha_{\rm \scalebox{0.5}{L}}}_{\rm \scalebox{0.5}{L}}\ket{\pm\alpha_{\rm \scalebox{0.5}{R}}}_{\rm \scalebox{0.5}{R}}$ is higher than that of the different-phase configuration $\ket{\pm\alpha_{\rm \scalebox{0.5}{L}}}_{\rm \scalebox{0.5}{L}}\ket{\mp\alpha_{\rm \scalebox{0.5}{R}}}_{\rm \scalebox{0.5}{R}}$. 
The correlation originates from the capacitive coupling between the JPOs, which corresponds to the coupling term in the Ising Hamiltonian shown in Eq.~(\ref{eq:Ising}), where the effective coupling is ferromagnetic for $\theta_{\rm p}=0$. 

\begin{figure}[h]
\centering
\includegraphics[width=\linewidth]{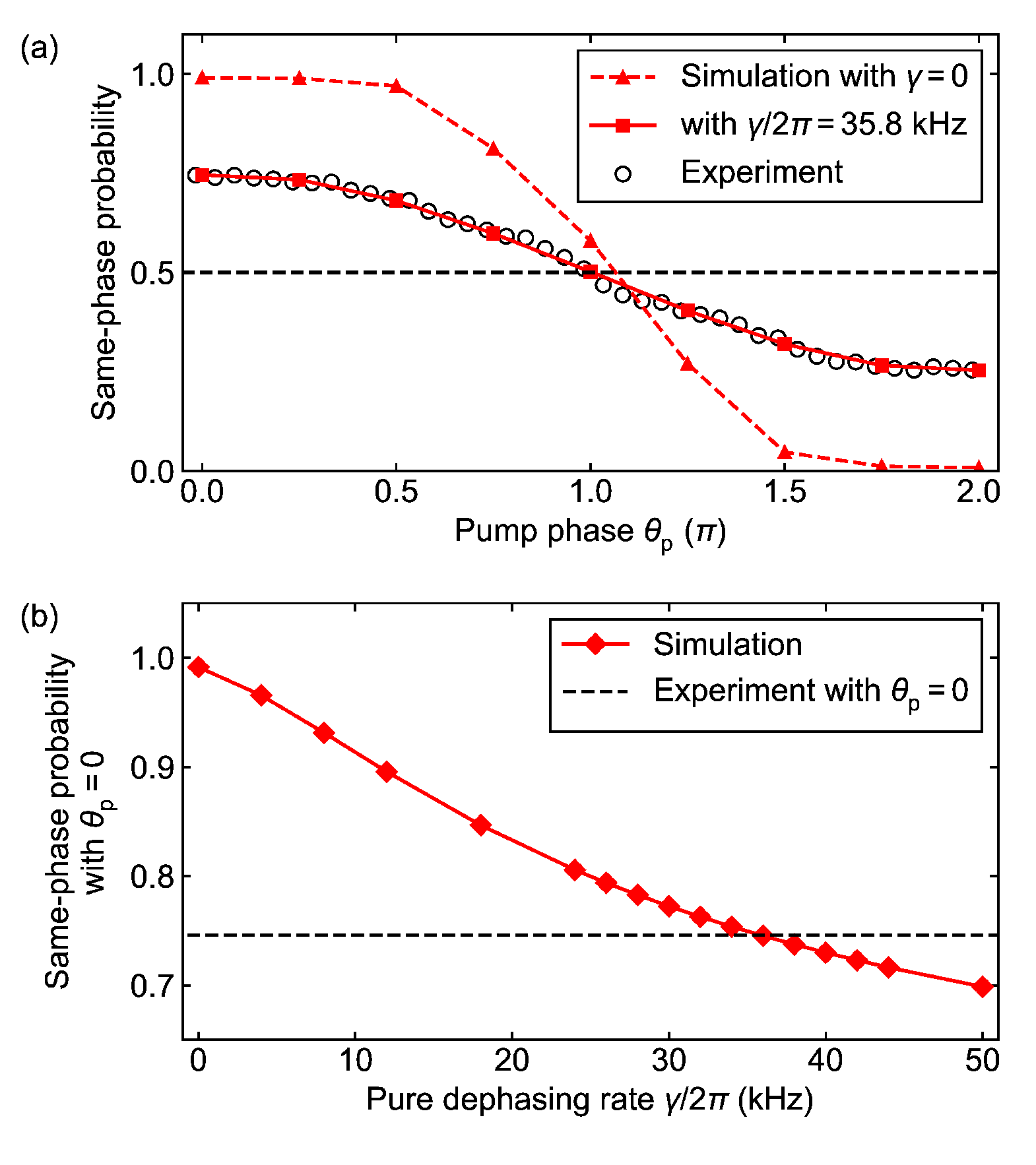}
\caption{
Relative-pump-phase and pure-dephasing-rate dependence of the correlation between the oscillation states of the two JPOs.
(a) Probability of the same-phase configuration as a function of the relative pump phase $\theta_{\rm p}$. 
The open circles show the experimental data, each of which is obtained by averaging $10^4$ shots. 
The red rectangles and the red triangles show the result of numerical simulation with the pure dephasing rate of $\gamma/2\pi=35.8$~kHz and 0~Hz, respectively, while the other parameters are set as those extracted from the experiments.  
(b) Probability of the same-phase configuration as a function of the pure dephasing rate $\gamma$. 
The red diamonds with the solid line show the numerical simulation with $\theta_{\rm p}=0$, while the other parameters are the same as (a). 
The black dashed line shows the probability of the experimental data at $\theta_{\rm p}=0$ shown in (a).
} 
 \label{fig:sim-exp-pphase}
\end{figure}

\begin{figure}[h]
\centering
\includegraphics[width=\linewidth]{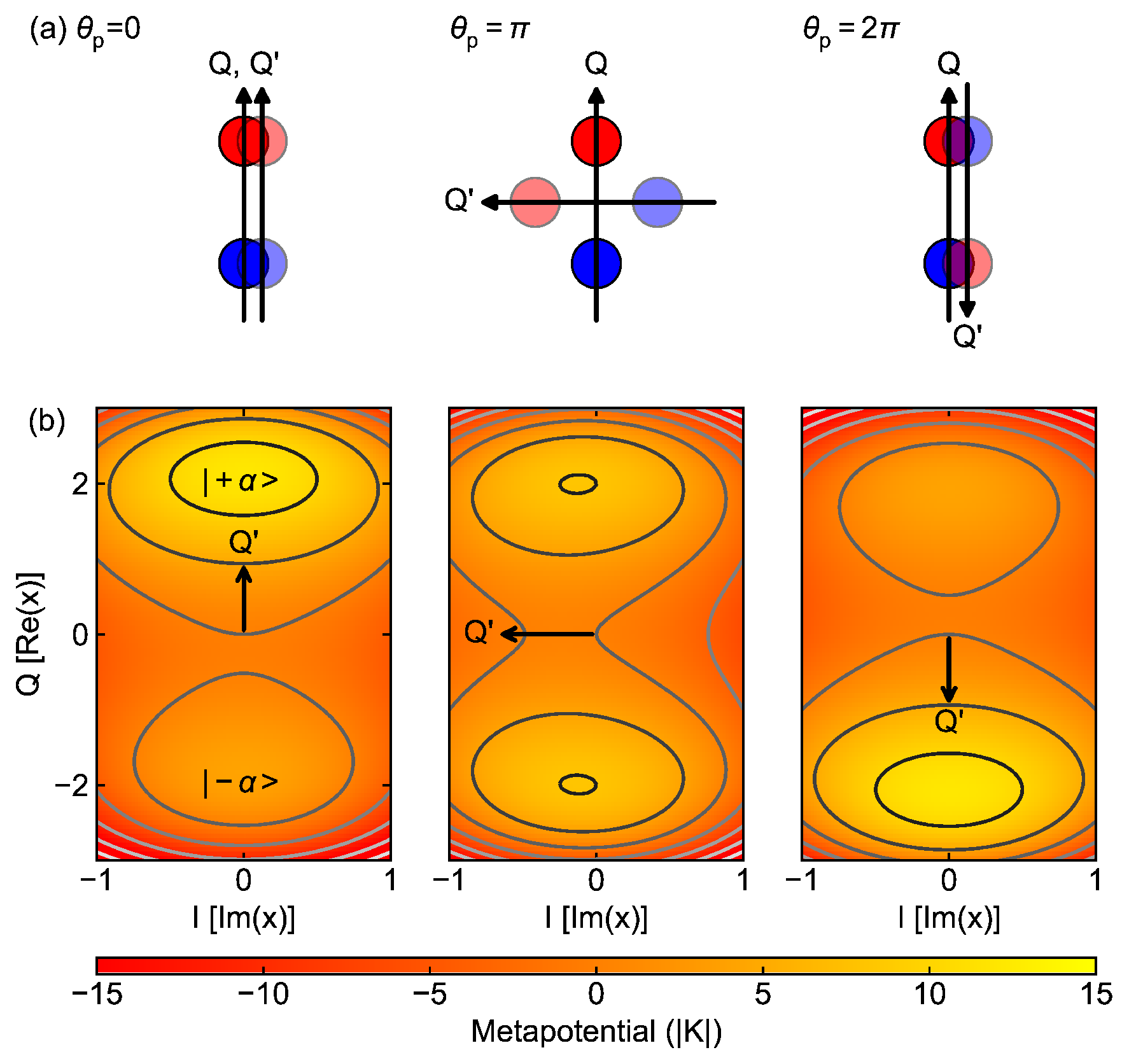}
\caption{
Effect of the relative pump phase. 
(a) Distribution of $\ket{\pm \alpha}$ states in the $IQ$-plane representation for $\theta_{\rm p}=0,\ \pi,$ and $2\pi$. 
The red and blue circles on the $Q$ axis show $\ket{+\alpha}$ and $\ket{-\alpha}$ states of the L JPO, respectively. 
The circles with paler colors show those of the R JPO, which are located on the $Q^\prime$ axis rotated by $\theta_{\rm p}/2$ due to the relative phase $\theta_{\rm p}$.
(b) Metapotential of the L JPO corresponding to (a) on the condition that the R JPO is $\ket{+\alpha}$ state. 
The color scale shows the metapotential of the L JPO, $\frac{K}{2} |x|^4 + \frac{p}{2}(x^{\ast 2} + x^2) + g(e^{-{\rm i}\theta_{\rm p}/2} \alpha_{\rm \scalebox{0.5}{R}} x^\ast + e^{{\rm i} \theta_{\rm p}/2} \alpha_{\rm \scalebox{0.5}{R}}^\ast  x)$ in units of $|K|$, where $p=4|K|$ and $g\alpha_{\rm \scalebox{0.5}{R}}=|K|$. 
The horizontal and vertical axes show $I$ $[{\rm Im}(x)]$ and $Q$ $[{\rm Re}(x)]$ amplitudes, respectively. 
The coupling inclines the metapotential along the $Q^\prime$ axis shown by the black arrows.
} 
 \label{fig:explanation_pphase}
\end{figure}

Equation~(\ref{eq:Ising}) predicts that the magnitude and sign of the effective coupling can be controlled by changing the relative pump phase $\theta_{\rm p}$, and here we confirm it by changing the R pump phase. As shown in Figs. \ref{fig:correlation}(d-e), the correlation disappears at $\theta_{\rm p}=\pi$ and becomes antiferromagnetic at $\theta_{\rm p}=2\pi$.
Figure~\ref{fig:sim-exp-pphase}(a) shows the occurrence probability of the same-phase configuration as a function of $\theta_p$.
The maximum probability is $75\%$, and the magnitude of the correlation has a cosine-shaped dependence on $\theta_{\rm p}$ in accordance with $J\propto {\rm cos}(\theta_{\rm p}/2)$.
The $\theta_{\rm p}$ dependence can be intuitively understood as follows. 
For JPO $i$, the coupling term in Eq.~(\ref{eq:H}) can be regarded as a coherent drive term by replacing $a_{j\ne i}$ with $\pm \alpha_j$.
Importantly this drive field depends on the state of JPO $j$, and thus it generates correlation between the JPOs. 
Also, the impact of the state-dependent coherent drive depends on $\theta_{\rm p}$ since the distribution of the JPOs on the $IQ$ plane rotates as shown in Fig.~\ref{fig:explanation_pphase}(a). 
Since the offset of the rotating frame is half the pump phase, the $IQ$ axes of the R JPO, the $I^\prime$ and $Q^\prime$ axes, are rotated by $\theta_{\rm p}/2$ relative to those of the L JPO. 
When the pump phases of the JPOs coincide with each other ($\theta_{\rm p}=0$), the $Q$ and $Q^\prime$ axes are parallel to each other, and the coupling inclines the metapotential of the L JPO along the $Q$ axis assuming that the R JPO is in $\ket{+\alpha}$ state as shown in the left panel of Fig.~\ref{fig:explanation_pphase}(b). 
The inclination induces the ferromagnetic correlation, which is maximal at $\theta_{\rm p}=0$.

The overlap between the $Q$ and $Q^\prime$ axes and the correlation decrease as $\theta_{\rm p}$ increases in the region of $0<\theta_{\rm p}<\pi$, and the effective coupling vanishes when $\theta_{\rm p}=\pi$, where the $Q$ axes are perpendicular to each other. 
The coupling at $\theta_{\rm p} = \pi$ inclines the metapotential along the $I$ axis and affects the metapotential around the $\ket{\pm \alpha}$ states equally [Fig.~\ref{fig:explanation_pphase}(b) center].
The correlation becomes antiferromagnetic in the region of $\pi<\theta_{\rm p} \le 2\pi$. 
The antiferromagnetic correlation is maximal at $\theta_{\rm p}=2\pi$, where the $Q$ and $Q^\prime$ axes are totally antiparallel, which corresponds to a reversal of the definition of the oscillation state.
 In this case, the metapotential of the L JPO is inclined along negative $Q$ direction [Fig.~\ref{fig:explanation_pphase}(b) right]. 
This characteristic of the metapotential shows the magnitude and polarity of the effective coupling can be easily controlled by varying the pump phases with a fixed capacitive coupling.

We compare the experimental result with numerical simulations taking into account pure dephasing rate as a fitting parameter (See Appendix~\ref{sup:numerical} for details).
In the numerical simulation, we assume that the pure dephasing rates of the JPOs are identical, that is, $\gamma_{\rm \scalebox{0.5}{L(R)}}=\gamma$, and the value of $\gamma$ is chosen so that the probability of the same-phase configuration agrees with the measured one for $\theta_p=0$. 
The experimental result is well reproduced by the simulation with $\gamma/2\pi=35.8~{\rm kHz}$, which is consistent with the upper limit from the spectroscopically measured internal photon loss rates $\gamma_{\rm \scalebox{0.5}{L(R)}}<\kappa_{\rm i\scalebox{0.5}{L(R)}}/2$.
We also investigate the pure-dephasing-rate dependence of the maximum correlation at $\theta_{\rm p}=0$ as shown in Fig.~\ref{fig:sim-exp-pphase}(b). 
The numerical simulation shows that the occurrence probability of the same-phase configuration is a monotonically decreasing function of $\gamma$ and close to 100\% without pure dephasing $\gamma=0$ as shown in the red triangles with the dashed line shown in Fig.~\ref{fig:sim-exp-pphase}(b).  
The maximum correlation is reduced by pure dephasing because pure dephasing induces the bit flip of parametric oscillations \cite{Puri2017}.
The independent evaluation and the improvement of the pure dephasing rate will be an important topic for the future study. 

The probability of the same-phase configuration is larger than 0.5 at $\theta_{\rm p}=\pi$ for $\gamma=0$ possibly because photon loss orients the $Q$ axis of a JPO \cite{Puri2017}. 
Since the orientation depends on the parameters of the JPOs as well as photon loss rate, the $Q$ axes of the JPOs are not exactly perpendicular at $\theta_{\rm p}=\pi$ due to the parameter asymmetry between the JPOs. 
Although the orientation becomes negligible as the pump amplitude increases, the effect of the orientation remains in the correlation for $\gamma=0$.  
On the other hand, bit flip caused by pure dephasing smears the effect for $\gamma/2\pi=35.8$~kHz.

Finally, we discuss the prospect for applying the present result to the quantum annealing.
Since the present study focuses on the demonstration of the controllability of the coupling, the experimental parameters, such as the pump detuning and the pump slope, need to be optimized for solving an Ising problem via adiabatic quantum evolution.
Adiabatic quantum evolution requires that the initial vacuum state is the highest energy state at $p_i=0$ in the rotating frame because the solution of the Ising Hamiltonian also corresponds to the highest energy state when the pump is applied.
This requires that $\Delta < -g < 0$ as shown in Appendix~\ref{sup:adiabatic}.
Since adiabatic quantum evolution can be faster than the time required for the system to reach a steady state, it can reduce the effect of pure dephasing on the probability of acquiring correct answers.
If the pumps are sufficiently detuned and a shorter pulse sequence is used, the maximum of the correlation can be larger than 90\% (See Appendix~\ref{sup:adiabatic} for details).
Since the negative $\Delta$ reduces the oscillation amplitude $\alpha_i \simeq \sqrt{(p_i + \Delta)/|K_i|}$ and the pump amplitude $p_i$ has an upper limit due to the critical current of the SQUID, shortening integration time may require the improvement of the signal-to-noise ratio (SNR).

Although the controllable two-body coupling scheme demonstrated in the present paper is directly applicable only to the network without loop structure, it can easily be extended to more general N-body interactions, {\it e.g.} four-body interaction used in the LHZ scheme \cite{Lechner2015}, which realizes scalable embedding of all-to-all connectivity. 

\section{conclusion}
We have designed and fabricated the device, where the two JPOs in the single-photon Kerr regime ($|K_i|\sim 10\kappa_{{\rm a}i}$) are capacitively coupled. 
We observed the simultaneous parametric oscillation by making the uncoupled resonance frequency of each JPO equal and simultaneously applying the pump at the same frequency.
The correlation between the oscillation phases of the JPOs up to 75\% is observed, and the amplitude and the polarity of the correlation can be controlled by varying the difference in the pump phases.
The tunability of the effective coupling is indispensable when mapping various optimization problems to Ising Hamiltonians and solving them with quantum annealing.
We have also simulated the pump-phase dependence by taking into account pure dephasing rate as an adjustable parameter. 
The experimental results are well reproduced by the simulation, which shows the validity of our experiment.
This experiment demonstrates the tunability of the Hamiltonian parameters, the coupling strength in the present case, by the phase of external microwave, which can be used in the Ising machine hardware composed of the KPO network. 

Various tradeoffs between the system parameters should be considered to maximize the performance of the quantum annealing. 
Since pure dephasing causes the bit flip of parametric oscillations, the shorter the time required to excite and measure parametric oscillations, the higher the probability of acquiring correct answers.
The parametric oscillations can be excited in a shorter time via an adiabatic evolution, which requires $\Delta < -g < 0$ in order to make the vacuum state the highest energy state in the rotating frame. 
The integration time can be shortened by increasing the external photon loss rate $\kappa_{{\rm e}i}$ ($<|K_i|$) and oscillation amplitude. 
However, a large $|K_i|$ and a negative $\Delta$ reduce oscillation amplitudes and the SNR.
The quantitative formulation of the optimal parameter is an important topic for the future study.

Although not used in the present paper, the local fields for the Ising Hamiltonian can be easily implemented by the phase-locking signal applied to each JPOs. 
Investigating the effect of local fields is also important in the application point of view. 
As pointed out in Ref.~\cite{Goto2018}, the output probability distribution of the KPO network obeys a Boltzmann distribution in the presence of dissipation due to a heating process called quantum heating \cite{Dykman2012, Dykman2011, Ong2013}.
Verifying the Boltzmann distribution of output phases with variable local fields is an interesting topic for the future study.

\begin{acknowledgments}
We thank Y. Kitagawa for his assistance in the device fabrication. 
We also thank Y.~Matsuzaki and K.~Matsumoto for useful discussions.
A part of this work was conducted at the AIST Nano-Processing Facility supported by ``Nanotechnology Platform Program'' of the Ministry of Education, Culture, Sports, Science and Technology (MEXT), Japan. 
The devices were fabricated in the Superconducting Quantum Circuit Fabrication Facility (Qufab) in National Institute of Advanced Industrial Science and Technology (AIST). 
This paper is based on results obtained from a project, JPNP16007, commissioned by the New Energy and Industrial Technology Development Organization (NEDO).

\end{acknowledgments}

\appendix

\section{device fabrication}
The device was fabricated on a high-resistive $380\ {\rm \mu m}$-thick silicon substrate. All the structures except Josephson junctions and airbridges were made of 100~nm-thick sputtered Nb film, which was dry etched using ${\rm CF_4}$ gas. 
The Josephson junctions were fabricated in a separate lithography step by shadow evaporations of Al, which was preceded by Ar-ion milling to remove the surface oxides of the Nb film. 
After the fabrication of the Josephson junctions, a positive photoresist was spin coated and the contact pad of the airbridges were defined by photolithography. 
We deposited 600~nm-thick sputtered Al film on the photoresist. We masked the bridge pattern by an additional positive photoresist, and wet etched the Al layer except the airbridges. Finally, we performed ${\rm O}_2$ ashing and removed all the photoresist using an NMP-based photoresist stripper.

\section{measurement setup}
\label{sup:meas-setup}
 \begin{figure}
	\includegraphics[width=\linewidth]{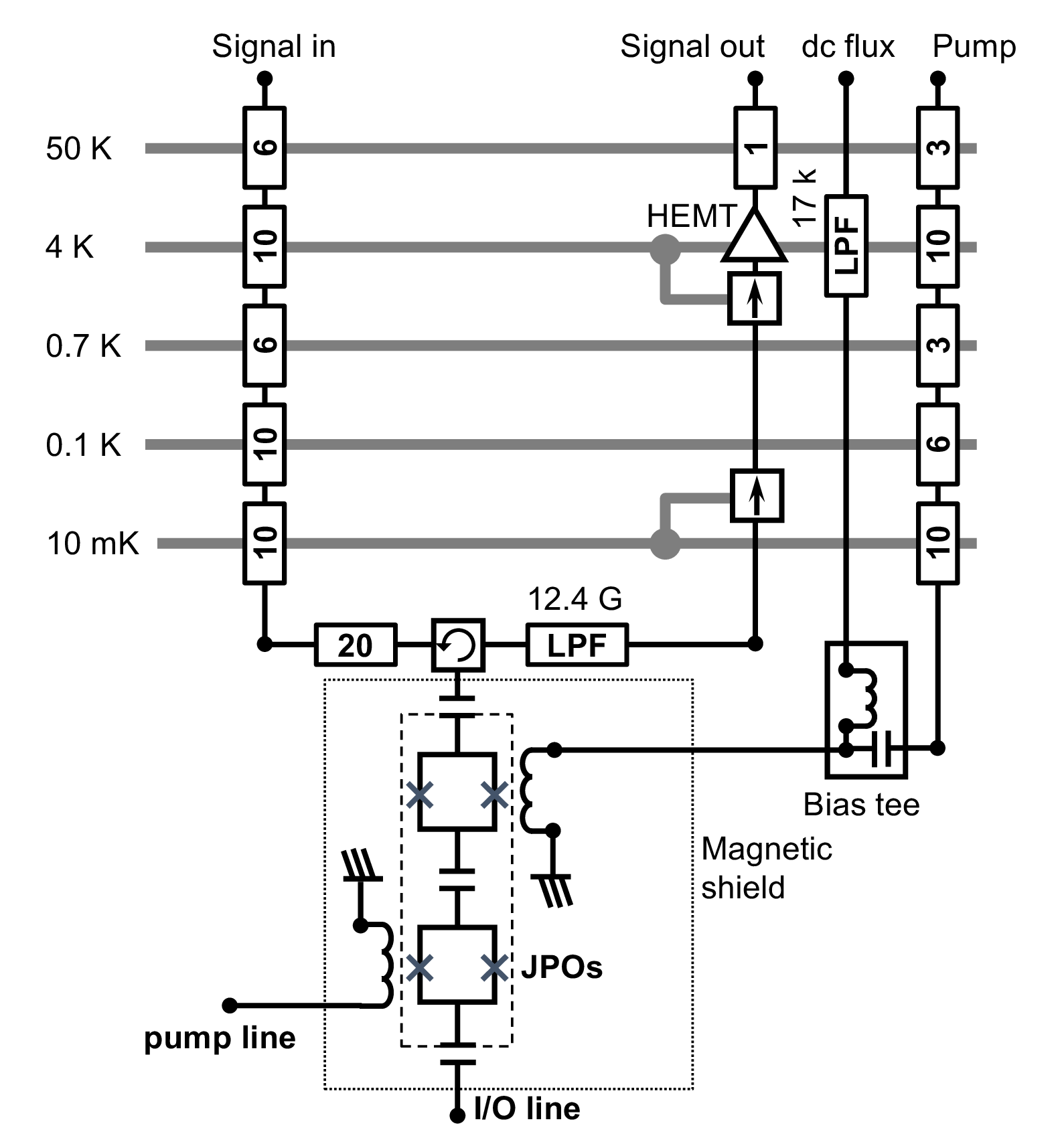}
	\caption{
	Measurement setup in the dilution refrigerator. 
	The horizontal gray lines show the temperature stages of the fridge. 
	The I/O and pump lines for only one of the two JPOs are shown.
	The rectangles with a number inside represent cryogenic attenuators with corresponding attenuation in decibels. 
	The rectangles with circular and straight arrows inside represent a circulator and an isolator, respectively. 
	Low-pass filters (LPFs) are shown with their cut-off frequency. 
	}
	\label{fig:rf_fridge}
\end{figure}

Figure~\ref{fig:rf_fridge} shows the measurement setup of the circuits in the dilution refrigerator. 
The device chip is installed inside the magnetic shield at the mixing chamber cooled below 10~mK. 
The input and pump lines are $50\Omega$ coaxial cables equipped with attenuators thermally anchored to each temperature stages of the refrigerator to reduce thermal noises from the room temperature environment. 
The total attenuation are 42~dB and 32~dB for the input and pump lines, respectively. 
The pump line is combined with the DC current line by a bias tee connected to the pump port of the JPO.
Probe microwave is injected into the sample via a circulator to route the reflection to the output line. 
The output is filtered by a low-pass filter, and then amplified by a cryogenic high electron mobility transistor amplifier.

 \begin{figure*}
	\includegraphics[width=\linewidth]{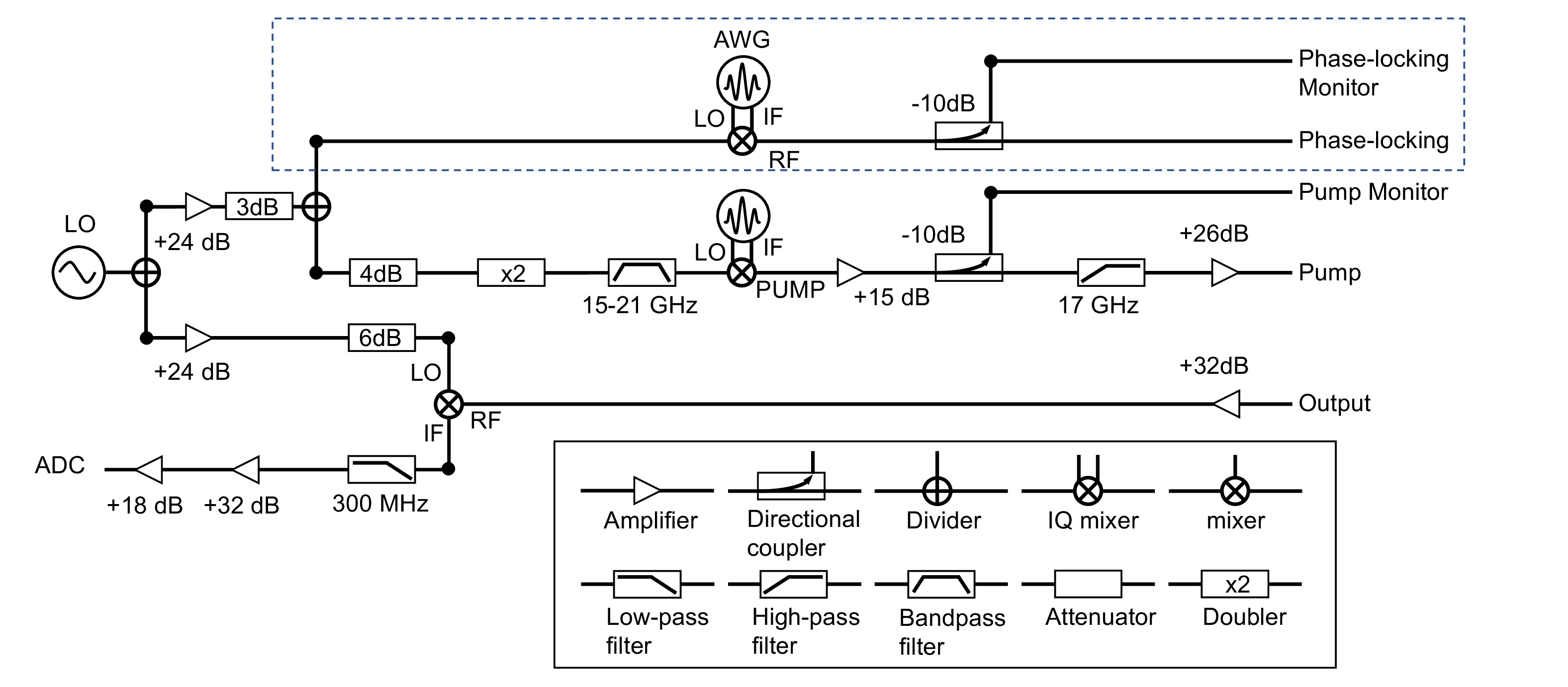}
	\caption{
	Setup of room-temperature electronics for the time-domain measurement. 
	The setup for one JPO is shown. 
	The dotted rectangle shows the line for the generation of phase-locking signals with a frequency of $\omega_{\rm p}/2$, which is not used in this paper.
	}
	\label{fig:rf_rt}
\end{figure*}

Figure~\ref{fig:rf_rt} shows the setup of the room-temperature electronics for the time-domain measurement. 
A Local oscillator (LO), Keysight M9347A DDS, provides the system with CW microwaves with a frequency of $(\omega_{\rm p}/2+\omega_{\rm \scalebox{0.5}{IF}}/2)/2\pi=10.3542$~GHz, where $\omega_{\rm \scalebox{0.5}{IF}}/2\pi=40$~MHz is the intermediate frequency. 
The microwaves are divided and supplied to the three mixers: two IQ mixers for the generation of the phase-locking signal and the pump and one mixer for the demodulation of the output signal from the device. 
For the generation of the phase-locking signal and the pump, we used arbitrary waveform generators, Keysight M3202A, with a sampling rate of 1~GSa/s for the baseband signal, although we did not use the phase-locking signal in the present paper. 
For the pump pulse, a frequency doubler is used to generate CW microwave at $(\omega_{\rm p}+\omega_{\rm \scalebox{0.5}{IF}})/2\pi=20.7084$~GHz used as an LO for the IQ mixer. 
The image sideband and the carrier frequency leakage are suppressed by more than $60$~dB by performing the calibration prior to the experiments. 
The outputs from the refrigerator are amplified by a room temperature amplifier and then down-converted to the intermediate frequency $\omega_{\rm IF}/2$ by the frequency mixer for the demodulation, and then recorded by an ADC, Keysight M3102A with a sampling rate of 500~MSa/s.

\section{Numerical method}
\label{sup:numerical}
In order to simulate the measurement, we numerically solve the master equation,
\begin{eqnarray}
\frac{d\rho(t)}{dt} &=& -\frac{i}{\hbar}[\mathcal{H}(t),\rho(t)] + \mathcal{L}[\rho(t)],\nonumber\\
\mathcal{L}[\rho] &=& \sum_{i={\rm L,R}} \frac{\kappa_{{\rm a}i}}{2} ([a_i\rho, a_i^\dagger] + [a_i,\rho a_i^\dagger])\nonumber\\
&&+\gamma_i ([a_i^\dagger a_i\rho, a_i^\dagger a_i] + [a_i^\dagger a_i,\rho a_i^\dagger a_i]), 
\label{ME_11_3_22}
\end{eqnarray}
where $\mathcal{H}(t)$ is the Hamiltonian shown in Eq.~(\ref{eq:H}) with a time-dependent pump amplitude; $\rho (t)$ is the density matrix, and $\gamma_i$ is the pure dephasing rate.
Here, $p_{\rm \scalebox{0.5}{L(R)}}$ is chosen so that $\alpha_{\rm \scalebox{0.5}{L(R)}}=\sqrt{p_{\rm \scalebox{0.5}{L(R)}}/|K_{\rm L(R)}|}$ coincides with the measured ones, 2.8 (2.5).
As mentioned in the main text, the measured internal loss rates $\kappa_{{\rm i}i}$ include the contribution from the pure dephasing.
The actual internal loss rate, $\kappa_{{\rm i}i}^\ast$, is related to the measured one as $\kappa_{{\rm i}i}^\ast= \kappa_{{\rm i}i} - 2\gamma_i$, and the total photon loss rate $\kappa_{{\rm a}i}$ in Eq.~(\ref{ME_11_3_22}) is defined as $\kappa_{{\rm a}i}\equiv \kappa_{{\rm e}i} + \kappa_{{\rm i}i}^\ast$.
All the parameters except $\gamma_i$ are determined from the measurements. 
We assume that the JPOs are in the vacuum state at the initial time.
We define the amplitude of the same-phase configuration at time $t$ as 
\begin{eqnarray}
\xi_{+}(t)&=&\iint_{{\alpha_{\rm \scalebox{0.4}{L}}}{\alpha_{\rm \scalebox{0.4}{R}}}>0} d{\alpha_{\rm \scalebox{0.5}{L}}}d{\alpha_{\rm \scalebox{0.5}{R}}} \nonumber \\
&& \times (\alpha_{\rm \scalebox{0.5}{L}}+\alpha_{\rm \scalebox{0.5}{R}}) \big{|} \langle\alpha_{\rm \scalebox{0.5}{L}},\alpha_{\rm \scalebox{0.5}{R}}|\rho(t)|\alpha_{\rm \scalebox{0.5}{L}}, \alpha_{\rm \scalebox{0.5}{R}}\rangle \big{|},
\end{eqnarray}
where $\alpha_i$ is real, and the subscript of the integral indicates that the integration is performed in the regions where the product of $\alpha_{\rm \scalebox{0.5}{L}}$ and $\alpha_{\rm \scalebox{0.5}{R}}$ is positive.
The amplitude of the different-phase configuration $\xi_{-}$ is defined in an analogous manner.
The probability of the same-phase configuration at time $t$ is defined by
\begin{eqnarray}
p'_+(t) = \frac{\xi_{+}(t)}{\xi_{+}(t)+\xi_{-}(t)}.
\label{p_p_11_5_22}
\end{eqnarray}
This probability is averaged during the time corresponding to the readout.
We refer to the time-averaged probability as the correlation of the same-phase configuration.

\section{Optimization for quantum annealing}
\label{sup:adiabatic}
\begin{figure}[h]
\centering
\includegraphics[width=\linewidth]{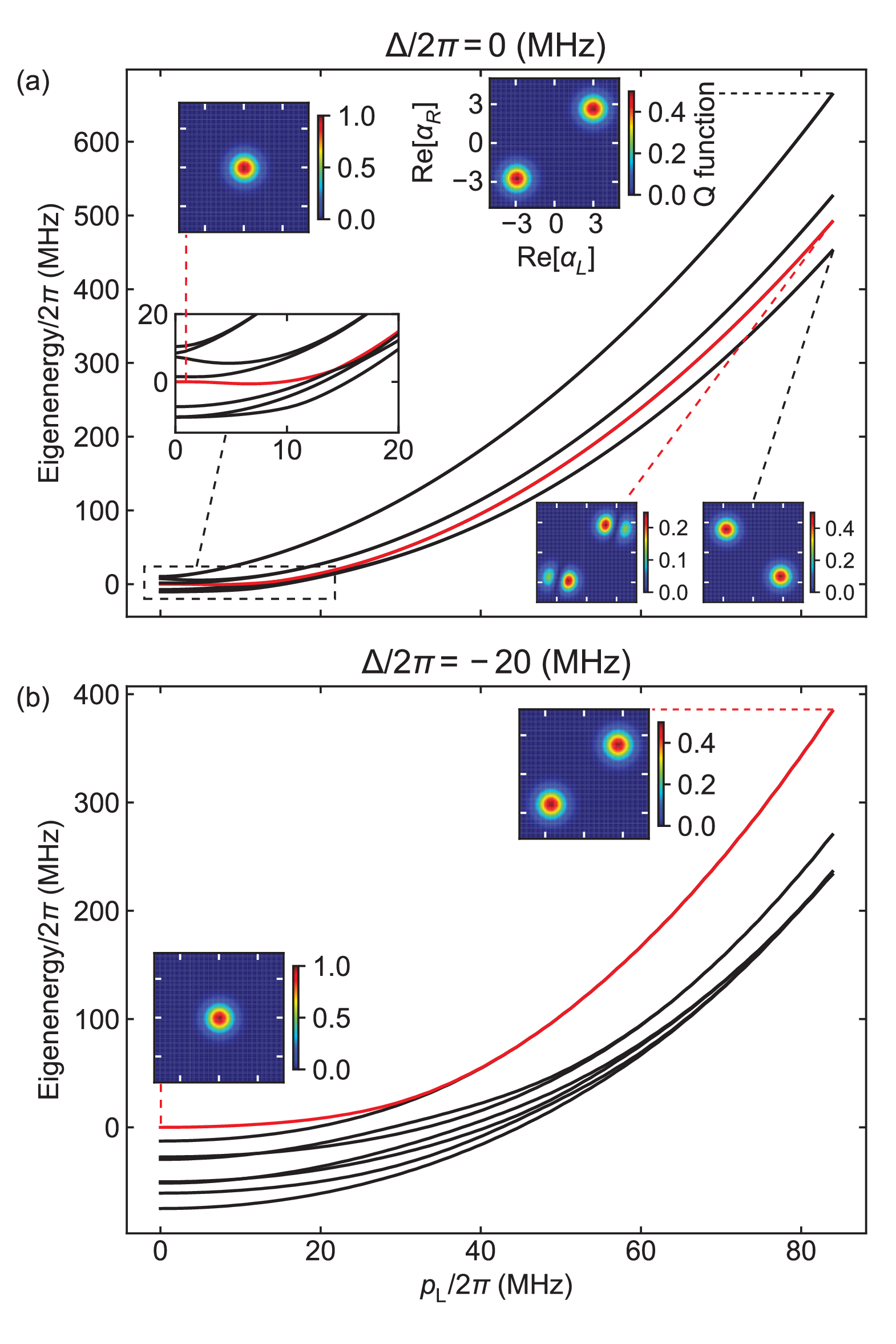}
\caption{
The pump-amplitude dependence of the eigenenergies for (a) $\Delta=0$ and (b) $\Delta/2\pi=-20$~MHz. 
The horizontal axis shows the pump amplitude $p_{\rm \scalebox{0.5}{L}}$, where we set the ratio between $p_{\rm \scalebox{0.5}{L}}$ and $p_{\rm \scalebox{0.5}{R}}$ to be the same as the experiment shown in Figs.~\ref{fig:correlation} and \ref{fig:sim-exp-pphase}.
The vertical axis represents the eigenenergies of the Hamiltonian shown in Eq.~(\ref{eq:H}), where $\theta_{\rm p}$=0 and other parameters are based on the experiment shown in Fig.~\ref{fig:correlation}(c). 
The eight highest eigenenergies are plotted. 
The solid red line is for the energy eigenstate which coincides with the vacuum state at $p_{\rm \scalebox{0.5}{L(R)}}=0$. 
Husimi Q functions $\bra{\alpha}\rho\ket{\alpha}$ for the corresponding states indicated by the dashed lines are plotted as a function of the real part of $\alpha_{\rm \scalebox{0.5}{L}}$ (horizontal axis) and that of $\alpha_{\rm \scalebox{0.5}{R}}$ (vertical axis).
All the scales of the horizontal and vertical axes of the Q functions are the same.
}
 \label{fig:sim-energy}
\end{figure}

\begin{figure}[h]
\centering
\includegraphics[width=\linewidth]{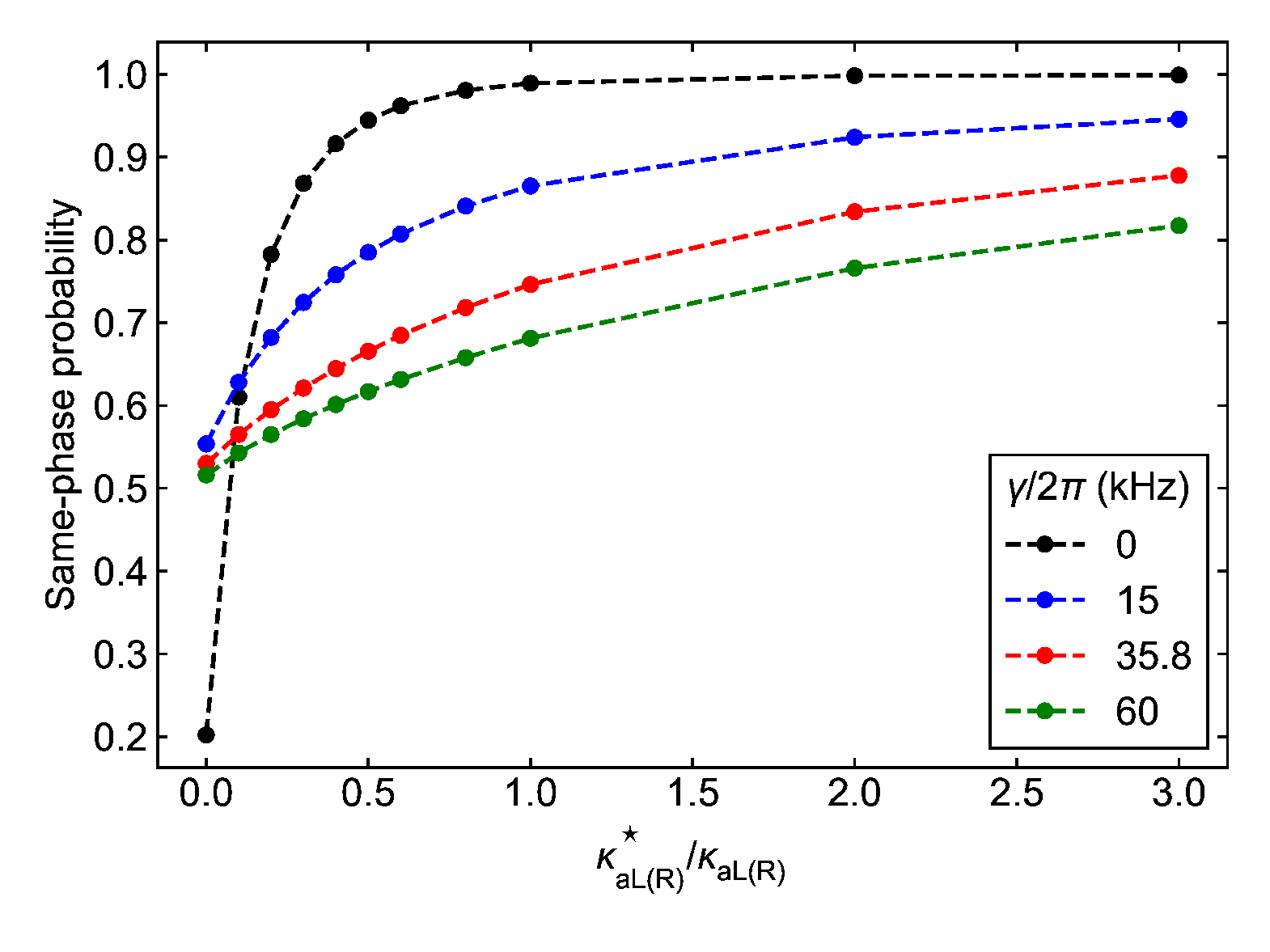}
\caption{
The $\kappa$ dependence of the simulated occurrence probability of the same-phase configuration. 
The black, blue, red, and green circles are for the pure dephasing rate of $\gamma/2\pi=0$, 15, 35.8, and 60~kHz, respectively.
The horizontal axis shows the total photon loss rates of the JPOs normalized by those obtained from the experiment. 
We set the total photon loss rates of the JPOs in the simulation as $\kappa_{{\rm a} i}^\star=M\kappa_{{\rm a} i}$, where $\kappa_{\rm a\scalebox{0.5}{L(R)}}/2\pi=1.00(0.86)$~MHz and $0 \le M \le 3$.
The other parameters are the same as Figs.~\ref{fig:correlation}(c) and \ref{fig:sim-exp-pphase}(a) (the experimental data at $\theta_{\rm p}=0$).  
}
 \label{fig:sim-kappa}
\end{figure}

Figure~\ref{fig:sim-energy} shows the pump-amplitude dependence of the eigenenergies of the system.
When $\Delta=0$, which is the case in our experiments, the vacuum state is not the highest energy state for $p_{\rm \scalebox{0.5}{L(R)}}=0$ due to the coupling as shown in Fig.~\ref{fig:sim-energy}(a).  
When the pump amplitude is ramped up, photon loss causes the transition of the system from the vacuum state, whose Q function is shown in the upper left inset, to the highest energy state corresponding to the solution of the Ising model defined by Eq.~(\ref{eq:Ising}) (the upper right inset). 
The 2D plot of the Q function for the vacuum state, $\ket{0}_{\rm \scalebox{0.5}{L}}\ket{0}_{\rm \scalebox{0.5}{R}}$, is localized near the origin. 
In contrast, the Q function for the state corresponding to the solutions of the Ising model, which is approximately a superposition of $\ket{\pm\alpha_{\rm \scalebox{0.5}{L}}}_{\rm \scalebox{0.5}{L}}\ket{\pm\alpha_{\rm \scalebox{0.5}{R}}}_{\rm \scalebox{0.5}{R}}$, shows two peaks corresponding to the same-phase configurations.
Figure~\ref{fig:sim-kappa} shows the $\kappa$ dependence of the simulated occurrence probability of the same-phase configuration.
The correlation is a monotonically increasing function of the photon loss rate, indicating that the transition to the correlated parametric oscillations is due to the relaxation caused by photon loss. 
We note that when the pure dephasing rate and the photon loss rate are zero, namely without decoherence, the same-phase probability is approximately 0.2. 
Numerical simulation shows that the final population of one of the lowest degenerate states shown in Fig.~\ref{fig:sim-energy}(a), which has even parity and the opposite correlation [the bottom rightmost inset in Fig.~\ref{fig:sim-energy}(a)], is higher than 65\%. 
This indicates that there are nonadiabatic population transfers presumably due to the narrow gaps between levels at around $p/2\pi$=15 MHz.

\begin{figure}[h]
\centering
\includegraphics[width=\linewidth]{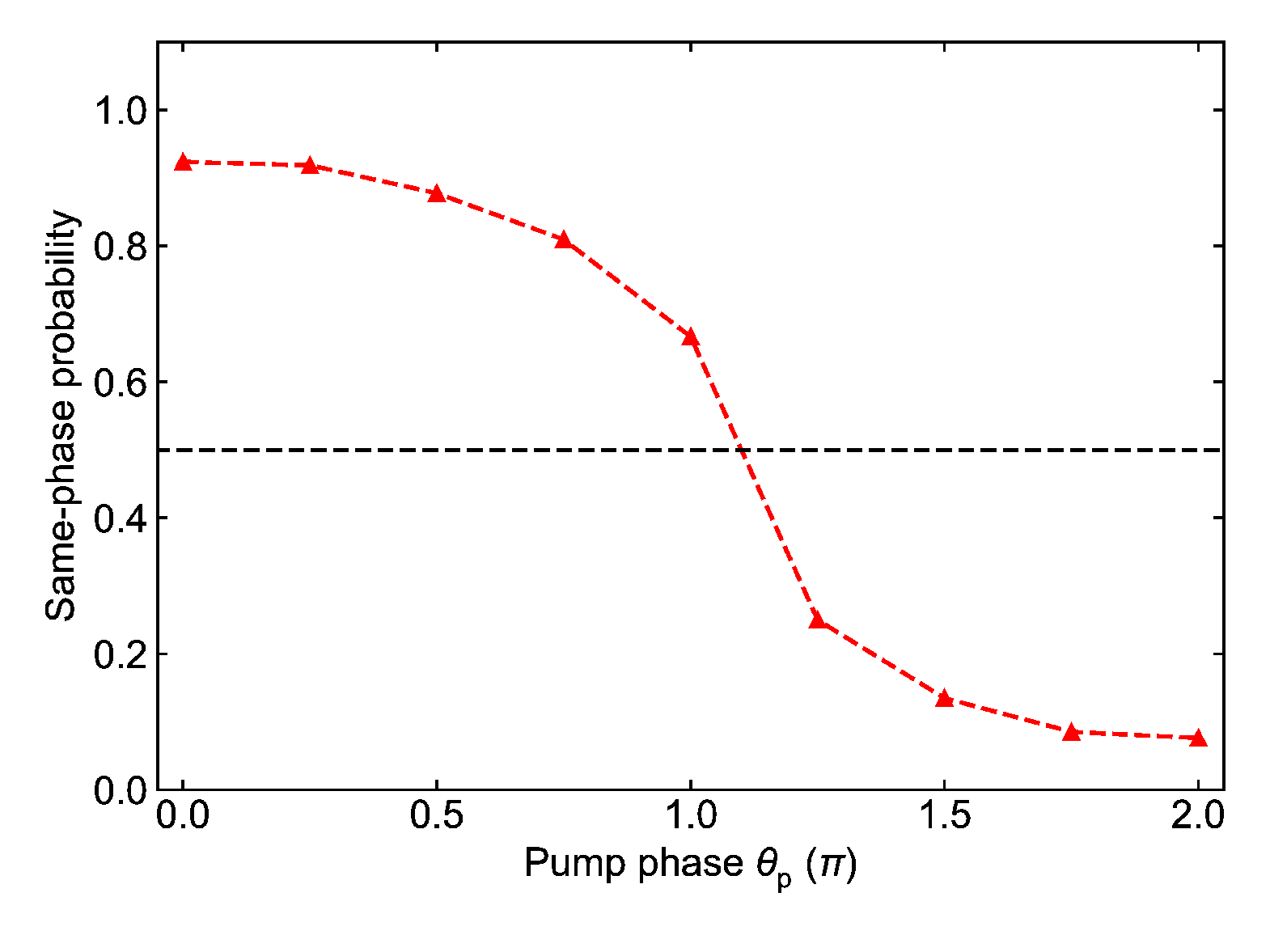}
\caption{
The simulated $\theta_{\rm p}$ dependence of the correlation with $\Delta/2\pi=-20$~MHz and a shorter pulse sequence. 
The vertical axis shows the occurrence probability of the same-phase configuration.
The pulse sequence is set as follows: pump slope is 100~ns, there is no delay in readout from the start of the plateau of the pump, and the readout time is 200~ns, which may require the improvement of measurement SNR by using a Josephson parametric amplifier \cite{Aumentado2020}, for example. 
The other parameters are the same as the simulation with $\gamma/2\pi=35.8$~kHz shown in Fig.~\ref{fig:sim-exp-pphase}(a).  
}
 \label{fig:sim-new-scheme}
\end{figure}

The energy levels of the states other than the vacuum state can be lowered by a negative detuning $\Delta<0$, and the vacuum state becomes the highest energy level at $p_{\rm \scalebox{0.5}{L(R)}}=0$ when $\Delta<-g$ as exemplified in Fig.~\ref{fig:sim-energy}~(b).  
In this case, the system remains at the highest level during the increase of the pump amplitude. 
The increase of the pump amplitude should be slow enough to suppress unwanted nonadiabatic transitions and fast enough to avoid the effect of photon loss. 
Then, the system evolves from the vacuum state [the left inset in Fig.~\ref{fig:sim-energy}~(b)] to the state corresponding to the solution of the Ising model (the right inset).
These coherent dynamics are expected to improve the occurrence probability of the solution of the Ising model compared to the one based on the system's relaxation caused by photon loss because the shorter pulse sequence reduces the effect of pure dephasing.
Figure~\ref{fig:sim-new-scheme} shows the $\theta_{\rm p}$ dependence of the correlation on the condition that $\Delta=-20~{\rm MHz}<-g$ and the pump slope is reduced to 100~ns, which is larger than $1/K$ and long enough to avoid unwanted nonadiabatic transitions.
Since the negative detuning 
enables the adiabatic evolution from the initial vacuum state to the correlated parametric oscillations before the system reaches the steady state, the maximum correlation is improved to $>$90\%.

\bibliography{master}

\end{document}